\documentclass{elsart}
\usepackage{graphicx,amsmath,amssymb,bm}

\begin{document}

\def\pdag{p\hspace{-1.8mm} / }
\def\tpdag{p\hspace{-1.3mm} / }
\def\qdag{q\hspace{-2.0mm} / }
\def\tqdag{q\hspace{-1.4mm} / }
\def\kdag{k\hspace{-2.2mm} / }
\def\tkdag{k\hspace{-1.5mm} / }

\begin{frontmatter}

\title{Magnetic dipole moment of the $\bm{S_{11}}$(1535)
       from the $\bm{\gamma p \to \gamma \eta p}$ reaction}

\author{Wen-Tai Chiang and Shin Nan Yang}
\address{Department of Physics, National Taiwan University,
         Taipei 10617, Taiwan}
\author{Marc Vanderhaeghen and Dieter Drechsel}
\address{Institut f\"ur Kernphysik, Universit\"at Mainz,
         55099 Mainz, Germany}


\begin{abstract}
The $\gamma p \to \gamma \eta p$ reaction in the $S_{11}$(1535) resonance
region is investigated as a method to access the $S_{11}$(1535) magnetic
dipole moment. To study the feasibility, we perform calculations of the
$\gamma p \to \gamma \eta p$ process within an effective Lagrangian
approach containing both the $S_{11}$ resonant mechanism and a background
of non-resonant contributions. Predictions are made for the forthcoming
experiments. In particular, we focus on the sensitivity of cross sections
and photon asymmetries to the $S_{11}$(1535) magnetic dipole moment.
\end{abstract}


\end{frontmatter}

\section{Introduction}
\label{sec:Intro}%
The study of the properties of the $S_{11}(1535)$ $N^*$ resonance, the
lowest lying $J^P = {\frac{1}{2}}^-$ nucleon resonance, provides valuable
insight into the nature of QCD in the non-perturbative domain. In
particular, the large mass splitting between the nucleon ground state
$N(939)$ and its negative parity partner $N^*(1535)$ is connected to the
spontaneous breaking of the chiral symmetry. Indeed, if the two-flavor
chiral symmetry were exact and preserved by the QCD vacuum, QCD would
predict parity doublets degenerate in mass~\cite{DeTar:1988kn,Jido:2001nt}.

In recent years, new techniques have become available to perform numerical
calculations of the properties of nucleon excited states using lattice QCD
\cite{Sasaki02,Gock02,Mel02}. In particular, calculations of the mass
spectrum of nucleon excited states were performed~\cite{Sasaki02}, which
implement chiral symmetry at non-zero lattice spacing, by use of domain
wall fermions. The recent quenched lattice QCD calculations
\cite{Sasaki02,Gock02,Mel02}, using different fermion actions, all
reproduce quite well the observed large mass splitting between the nucleon
and its negative parity partner $N^*(1535)$. Furthermore, the next
generation of lattice QCD calculations opens up the exciting prospect to
study how nucleon and resonance properties change when varying the quark
mass parameter from the large quark mass regime, where chiral symmetry is
explicitly broken, down to sufficiently small quark mass values, where one
can use the chiral symmetry of QCD to perform the extrapolation to the
chiral limit~\cite{Thomas02}. The large quark mass region may be the regime
where a naive constituent quark picture could hold, and one may investigate
how such a picture breaks down when approaching physical quark masses.

A particularly useful observable to study in this respect is the baryon
magnetic dipole moment (MDM), as it has often been questioned
and discussed \cite{Mor92,Cheng02,DM02} why the naive non-relativistic
constituent quark model yields a relatively
good agreement with the experimentally known values for the MDMs of the
ground state octet baryons. Recently, it has been studied
\cite{Leinweber99,Leinweber01} how the proton and neutron MDMs change when
extrapolating from large up- and down-quark masses (about 6 - 10 times
their physical mass values) where current lattice QCD calculations are
performed, down to small quark masses, where the non-analytic behavior in
the quark mass can be calculated from the chiral symmetry of QCD. In
particular, it has been noticed in Ref.~\cite{Leinweber01} that the
apparent success of the non-relativistic
constituent quark model in reproducing the observed
ratio of proton to neutron MDMs at the physical quark masses is rather
coincidental, supporting the finding of Ref.~\cite{Mor92}.
Therefore, it may be very revealing to extend such a study to
the MDMs of nucleon resonances, such as the $\Delta(1232)$ and the
$S_{11}(1535)$.

The $\Delta(1232)$ MDMs have been investigated on the lattice at rather
large quark masses \cite{Leinweber92}, and very recently the chiral
extrapolation of the $\Delta(1232)$ MDMs, including the next-to-leading
non-analytic variation with the quark mass, was also studied
\cite{Cloet02}. As there is a large gap in quark mass to bridge between the
present quenched lattice QCD calculations and the chiral limit, it would be
extremely helpful to know the resonance MDM for the physical quark mass
values, through experiment. Unfortunately, the experimental information on
the MDMs beyond the ground state baryon octet is very scarce. With the
notable exception of the $\Omega^-$ baryon, these higher nucleon resonances
decay strongly, and thus have too short lifetimes to measure their MDMs in
the conventional way through spin precession measurements.

For the $\Delta^{++}(1232)$, it was therefore proposed to measure its MDM
through the $\pi^+ p \to \gamma \pi^+ p$ reaction~\cite{Kon68}, and two
experiments were performed \cite{Nef78,Bos91}. Using different theoretical
analyses, the Particle Data Group (PDG)~\cite{Hagiwara:2002pw} quotes as
range of the MDM~: $\mu_{\Delta^{++}} = 3.7 - 7.5 \, \mu_N$, where $\mu_N$
is the nuclear magneton. The large uncertainty in the extraction of
$\mu_{\Delta^{++}}$ from the data is due to large non-resonant
contributions to the $\pi^+ p \to \gamma \pi^+ p$ reaction because of
bremsstrahlung from the charged pion ($\pi^+$) and proton (p).
Alternatively, it has been proposed~\cite{Dre83} to determine the magnetic
moment of the $\Delta^+(1232)$ through measurement of the $\gamma p \to
\gamma \pi^0 p$ reaction, thus avoiding bremsstrahlung contributions from
the pions. First calculations for this reaction, including besides the
$\Delta(1232)$ resonant mechanism~\cite{Mac99,DVGS} also a background of
non-resonant contributions, have recently been performed
~\cite{Drechsel:2001qu,Mach02}. Due to the small cross sections of the
$\gamma p \to \gamma \pi^0 p$ reaction, the first experimental data have
been obtained only very recently by the A2/TAPS Collaboration at the MAMI
accelerator~\cite{Kot99}. The pioneering experiment of Ref.~\cite{Kot99}
sees a clear deviation from the known soft bremsstrahlung processes (for
which the final photon is radiated from the external proton lines), and the
measured cross sections are in qualitatively good agreement with the
calculation of Ref.~\cite{Drechsel:2001qu}. The deviation from the soft
bremsstrahlung contains a sensitivity to the magnetic dipole strength at
the $\Delta(1232)$ resonance position. The experiment of Ref.~\cite{Kot99}
does not yet allow for a reliable extraction of this quantity, because of
limited statistics. However, a dedicated experiment to measure the $\gamma
p \to \gamma \pi^0 p$ reaction is planned at MAMI in the near future
\cite{CB}. This experiment will use the $4 \pi$ Crystal Ball detector and
thus increase the count rates by two orders of magnitude. In addition, this
forthcoming experiment ~\cite{CB} will be performed with a polarized photon
beam, because the sensitivity of the $\gamma p \to \gamma \pi^0 p$ reaction
to the $\Delta^+(1232)$ MDM can be increased substantially when using
linearly polarized photons, as has been shown in
Ref.~\cite{Drechsel:2001qu}.

In the present paper, we explore the extension of this technique to access
the MDM of the $S_{11}(1535)$ resonance. Since the $S_{11}(1535)$ resonance
is known to dominate the $\gamma p \to \eta p$ reaction in the threshold
region, the $\gamma p \to \gamma \eta p$ reaction, with a low energy
outgoing photon (up to about 100~MeV), seems to be a promising tool to
isolate the MDM of the $S_{11}(1535)$ resonance. Such an experiment can be
envisaged using the Crystal Barrel@ELSA and in the near future with the
Crystal Ball@MAMI, once the higher beam energy of about 1.5 GeV becomes
available at MAMI-C.

In Section~\ref{sec:modelmdm}, we provide some model estimates for the
$S_{11}(1535)$ MDM. In particular, we present the calculation of the
$S_{11}(1535)$ MDM in the constituent quark model and contrast it with the
value taken in the picture of the $S_{11}(1535)$ as a $(K \Sigma)$
quasi-bound (molecular) state, as has been proposed in Ref.~\cite{KSW95}.

We then develop in Section~\ref{sec:Model} an effective Lagrangian model
for the $\gamma p \to \gamma \eta p$ reaction in the $S_{11}(1535)$ region,
which is obtained by coupling a photon in a gauge invariant way to an
analogous model for the $\gamma p \to \eta p$ reaction. The model contains
both the $S_{11}(1535)$ resonance contribution and a background of
non-resonant processes.

In Section~\ref{sec:soft}, we discuss the model-independent link between
the $\gamma p \to \eta p$ process and the $\gamma p \to \gamma \eta p$
process in the limit where the outgoing photon energy approaches zero. We
present a low energy theorem, which provides a strong constraint for both
the experiment and the theoretical description of the $\gamma p \to \gamma
\eta p$ reaction.

In Section~\ref{sec:results}, we then use the developed effective
Lagrangian model as a tool to study the feasibility of an experiment to
measure the $\gamma p \to \gamma \eta p$ reaction. In particular, we
investigate the sensitivity of the $\gamma p \to \gamma \eta p$ cross
sections and photon asymmetries to the $S_{11}(1535)$ MDM. The final
section summarizes our findings.

\section{Model calculations for the $\bm{S_{11}(1535)}$ magnetic moment}
\label{sec:modelmdm}%
In this section, we calculate the results for the MDM of the
$S_{11}(1535)$ in the framework of the constituent quark model, and within
the picture where the $S_{11}(1535)$ is dynamically generated as a $(K
\Sigma)$ quasi-bound state, as first proposed in Ref.~\cite{KSW95}.

\subsection{Magnetic moments in the constituent quark model}
\label{sec:QM}%
In the nonrelativistic SU(6) constituent quark model, the lowest-lying
neg\-a\-tive-parity nucleon resonances are $|N^2P_{1/2}\rangle$ and
$|N^4P_{1/2}\rangle$, where the usual spectroscopic notations ${}^2P_{1/2}$
and ${}^4P_{1/2}$ are used to indicate their total quark spin $S=1/2,\,3/2$
($2S+1=2,\,4$), orbital angular momentum $L=1$ ($P$-wave), and total
angular momentum $J=1/2$. In contrast, the ground-state baryon octet (e.g.,
$N$, $\Lambda$, $\Sigma$) and decuplet (e.g., $\Delta$) have the same
$S$-wave spatial wavefunctions with $L=0$, and thus $J=S$. The
wavefunctions of the $|N^2P_{1/2} \rangle$ and $|N^4P_{1/2} \rangle$ states
are given explicitly as
\begin{eqnarray} \label{eq:N24}
  & & | N^2P_{1/2} \rangle
   = \frac{1}{\sqrt{2}} \sum_{m_l m_s} \langle\,1\,\tfrac{1}{2}\ m_l\,m_s
                              \,|\, \tfrac{1}{2}\, \tfrac{1}{2} \,\rangle
   \\
  & & \,\times\,\biggl\{ \psi^\rho_{1m_l} \Bigl[ \tfrac{1}{\sqrt{2}} \bigl(
           \chi^\lambda_{m_s}\,\phi^\rho +
           \chi^\rho_{m_s}\,\phi^\lambda \bigr) \Bigr]
         + \psi^\lambda_{1m_l} \Bigl[ \tfrac{1}{\sqrt{2}} \bigl(
           \chi^\rho_{m_s}\,\phi^\rho -
           \chi^\lambda_{m_s}\,\phi^\lambda \bigr) \Bigr] \biggr\} \,,
  \nonumber \\ [1ex]
  & &| N^4P_{1/2} \rangle
   = \frac{1}{\sqrt{2}} \sum_{m_l m_s} \langle\,1\,\tfrac{3}{2}\,m_l\ m_s
                              \,|\, \tfrac{1}{2}\, \tfrac{1}{2} \,\rangle
  \Bigl[\,\psi^\rho_{1m_l}\,\chi^s_{m_s}\,\phi^\rho +
        \,\psi^\lambda_{1m_l}\,\chi^s_{m_s}\,\phi^\lambda\,\Bigr] \,,
\end{eqnarray}
where $\psi$, $\chi$, and $\phi$ denote the spatial, spin, and flavor
wavefunctions. The superscripts $s$ or $\rho$ ($\lambda$) of these
wavefunctions indicate that they are totally symmetric among three quarks,
or odd (even) under the exchange of the first two quarks. Further details
can be found in Ref.~\cite{Isgur:1977ef}.

However, the observed lowest-lying negative-parity nucleon resonances are
the $S_{11}(1535)$ and $S_{11}(1650)$, obtained as configuration mixtures
of the $|N^2P_{1/2}\rangle$ and $|N^4P_{1/2}\rangle$ SU(6) states,
\begin{eqnarray}
  |\, S_{11}(1535) \,\rangle &=&
  |\, N^2P_{1/2} \,\rangle \cos\vartheta -
  |\, N^4P_{1/2} \,\rangle \sin\vartheta \,, \\ \nonumber
  |\, S_{11}(1650) \,\rangle &=&
  |\, N^2P_{1/2} \,\rangle \sin\vartheta +
  |\, N^4P_{1/2} \,\rangle \cos\vartheta \,,
\end{eqnarray}
where $\vartheta$ denotes the mixing angle.

In a constituent quark model, the magnetic dipole moments of $qqq$ baryons
consist of contributions from both quark spin and orbital angular momentum,
i.e. $\bm{\mu} = \bm{\mu^S} + \bm{\mu^L}$ with
\begin{eqnarray} 
  \bm{\mu^S}
  &=& \sum_i \bm{\mu^s}_i
   =  \sum_i \frac{Q_i}{m_i}\, \bm{s}_i \,,
\\ 
  \bm{\mu^L}
  &=& \sum_i \bm{\mu^l}_i
   =  \sum_i \frac{Q_i}{2m_i}\, \bm{l}_i \,,
\end{eqnarray}
where the index $i$ sums over three quarks.

In Appendix~\ref{append:qm}, we present the calculation of the
spin and orbital contributions to
the magnetic dipole moments for the states $|N^2P_{1/2}\rangle$ and
$|N^4P_{1/2}\rangle$ with $J_z=+1/2$. The results are given by~:
\begin{eqnarray} \label{eq:mu1}
  \mu(N^2P_{1/2}{}^+) &=&
  \langle\, N^2P_{1/2}{}^+ \,|\, \mu_z \,|\,
  N^2P_{1/2}{}^+ \,\rangle \nonumber\\ [0.5ex]
   &=& \tfrac{1}{9}\,(2\mu_u + \mu_d)
    =  \,\tfrac{1}{3}\; \mu_N \,,
\\ [1ex] \label{eq:mu2}
  \mu(N^2P_{1/2}{}^{\,0\,}) &=&
  \langle\, N^2P_{1/2}{}^{\,0\,} \,|\, \mu_z \,|\,
  N^2P_{1/2}{}^{\,0\,} \,\rangle \nonumber\\ [0.5ex]
   &=& \tfrac{1}{9}\,(2\mu_d + \mu_u)
    =  \ 0 \,,
\\ [1ex] \label{eq:mu3}
  \mu(N^4P_{1/2}{}^+) &=&
  \langle\, N^4P_{1/2}{}^+ \,|\, \mu_z \,|\,
  N^4P_{1/2}{}^+ \,\rangle \nonumber\\ [0.5ex]
   &=& \tfrac{4}{9}\,(2\mu_u + \mu_d)
    =  \,\tfrac{5}{3}\; \mu_N \,,
\\ [1ex] \label{eq:mu4}
  \mu(N^4P_{1/2}{}^{\,0\,}) &=&
  \langle\, N^4P_{1/2}{}^{\,0\,} \,|\, \mu_z \,|\,
  N^4P_{1/2}{}^{\,0\,} \,\rangle \nonumber\\ [0.5ex]
   &=& \tfrac{4}{9}\,(2\mu_d + \mu_u)
    =  -\tfrac{1}{3}\; \mu_N \,,
\end{eqnarray}
where $\mu_z = \mu^S_z + \mu^L_z$, and the superscripts + and 0 denote the
charge of the resonance state. Here we assume the constituent quarks as
Dirac point particles and approximated $m_u=m_d=\tfrac{1}{3}\,m_N$. Using
the relation $\mu=Q/2m$, we obtain $\mu_u=Q_u/2m_u=2\,\mu_N$ and
$\mu_d=Q_d/2m_d= -\mu_N$, which are used to obtain the values in
Eqs.~(\ref{eq:mu1})-(\ref{eq:mu4}).

In addition, there are the cross terms for $\mu^S_z$ which are obtained from
mixing the $|N^2P_{1/2}\rangle$ and $|N^4P_{1/2}\rangle$ states,
\begin{eqnarray} 
  & & \langle\, N^4P_{1/2}{}^+\,|\, \mu^S_z \,|\, N^2P_{1/2}{}^+\,\rangle
   =  \langle\, N^2P_{1/2}{}^+\,|\, \mu^S_z \,|\, N^4P_{1/2}{}^+\,\rangle
      \nonumber\\ [0.5ex]
  &=& \tfrac{4}{9}\,(\mu_u - \mu_d)
   =  \tfrac{4}{\;3\;}\; \mu_N \,,
\\ [1ex]
  & & \langle\, N^4P_{1/2}{}^{\,0\,}\,|\, \mu^S_z \,|\, N^2P_{1/2}{}^{\,0\,}\,\rangle
   =  \langle\, N^2P_{1/2}{}^{\,0\,}\,|\, \mu^S_z \,|\, N^4P_{1/2}{}^{\,0\,}\,\rangle
      \nonumber\\ [0.5ex]
  &=& \tfrac{4}{9}\,(\mu_d - \mu_u)
   =  - \tfrac{4}{3}\; \mu_N \,.
\end{eqnarray}
On the other hand, there are no cross terms for $\mu^L_z$ because
$|N^2P_{1/2}\rangle$ and $|N^4P_{1/2}\rangle$ have orthogonal quark spin
states which are not affected by $\mu^L_z$.

Now the magnetic moments of the $S_{11}(1535)$ and $S_{11}(1650)$
resonances can be expressed in terms of the magnetic moments of the
$|N^2P_{1/2}\rangle$ and $|N^4P_{1/2}\rangle$ states and the cross terms,
\begin{eqnarray} \label{eq:mus11a}
  \mu(S_{11}(1535))
      &=& \mu(N^2P_{1/2}) \cos^2\vartheta + \mu(N^4P_{1/2}) \sin^2\vartheta
          \\ \nonumber
      &-& 2\,\langle\, N^2P_{1/2} \,|\,\mu^S_z\,|\, N^4P_{1/2} \,\rangle
          \sin\vartheta \cos\vartheta \,,
  \nonumber \\ \label{eq:mus11b}
  \mu(S_{11}(1650))
      &=& \mu(N^2P_{1/2}) \sin^2\vartheta + \mu(N^4P_{1/2}) \cos^2\vartheta
          \\ \nonumber
      &+& 2\,\langle\, N^2P_{1/2} \,|\,\mu^S_z\,|\, N^4P_{1/2}\,\rangle
          \sin\vartheta \cos\vartheta  \,.
\end{eqnarray}

The value of the mixing angle $\vartheta$ depends on the quark interaction.
Assuming a hyperfine interaction between the quarks, Isgur and
Karl~\cite{Isgur:1977ef} predicted a mixing angle $\vartheta =
\tan^{-1}{(\sqrt{5}-1)/2} \simeq -31.7^\circ$, which is close to the
empirical mixing angle $\vartheta \simeq -32^\circ$ found in
Ref.~\cite{Hey:1975nc}. Using the value $\vartheta = -31.7^\circ$ in
Eqs.~(\ref{eq:mus11a}) and (\ref{eq:mus11b}), we obtain
\begin{eqnarray}
  \mu_{S_{11}^+(1535)} &=&  1.89\;\mu_N \quad \mathrm{and} \quad
  \mu_{S_{11}^0(1535)}  =  -1.28\;\mu_N \,,  \nonumber\\
  \mu_{S_{11}^+(1650)} &=&  0.11\;\mu_N \quad \mathrm{and} \quad
  \mu_{S_{11}^0(1650)}  =   0.95\;\mu_N \,,  \nonumber
\end{eqnarray}
which agrees with the result by the Genova group~\cite{Genova}. Very
similar results can also be obtained when the interactions of the $S_{11}$
resonances coupling to the $\pi N$ and $\eta N$ channels are explicitly
considered~\cite{Arima:wy}.

\subsection{Magnetic moment of $S_{11}(1535)$ as a
$(K \Sigma)$ quasi-bound state}%
It has been proposed in Ref.~\cite{KSW95} that the $S_{11}(1535)$ can be
considered as a $(K \Sigma)$ quasi-bound state of mesons and baryons. Using
an SU(3) effective chiral Lagrangian and iterating the derived
coupled-channel potentials to infinite orders, the $S_{11}(1535)$ resonance
can be generated dynamically. This finding has also been verified within
the chiral unitary approach~\cite{Nieves01,Inoue:2001ip}. Especially, in
Ref.~\cite{Inoue:2001ip} it is found that the coupling of the
$S_{11}(1535)$ to the $\eta N$ channel is nearly as large as the one to the
$K\Sigma$ channel.

In a simplified picture, we regard the $S_{11}(1535)$ as a $(K \Sigma)$
quasi-bound state. Then the $S_{11}(1535)$ charge states can be written
as~:
\begin{eqnarray}
  |\, S^+_{11}(1535) \,\rangle &\,=\,&
\sqrt{\frac{1}{3}} \, (K^+ \, \Sigma^0) \,+\,
\sqrt{\frac{2}{3}} \, (K^0 \, \Sigma^+) \, , \nonumber \\
  |\, S^0_{11}(1535) \,\rangle &\,=\,&
\sqrt{\frac{2}{3}} \, (K^+ \, \Sigma^-) \,-\,
\sqrt{\frac{1}{3}} \, (K^0 \, \Sigma^0) \, . \label{eq:ksigma}
\end{eqnarray}
From Eq.~(\ref{eq:ksigma}), it then follows that the magnetic moment of
both $S_{11}(1535)$ charge states can be expressed in terms of the hyperon
magnetic moments,
\begin{eqnarray}
  \mu_{S_{11}^+(1535)} &\,=\,&
\frac{1}{3} \, \mu_{\Sigma^0} \,+\, \frac{2}{3} \, \mu_{\Sigma^+} \, ,
\nonumber \\
  \mu_{S_{11}^0(1535)} &\,=\,&
\frac{2}{3} \, \mu_{\Sigma^-} \,+\, \frac{1}{3} \, \mu_{\Sigma^0} \, .
\end{eqnarray}
Using the experimental values for the $\Sigma^+$ and $\Sigma^-$ magnetic
moments, and approximating the unknown $\Sigma^0$ magnetic moment by the
average of the $\Sigma^+$ and $\Sigma^-$ magnetic moments, one obtains for
the $S_{11}(1535)$ magnetic moments in the $(K \Sigma)$ molecule picture
the values~:
\begin{eqnarray}
  \mu_{S_{11}^+(1535)} &\,=\,& 1.86\;\mu_N \, , \nonumber \\
  \mu_{S_{11}^0(1535)} &\,=\,& -0.56\;\mu_N \, .
\end{eqnarray}

We like to stress that the calculation given here for the MDM of the
$S_{11}(1535)$ as a $(K \Sigma)$ quasi-bound state is only a very simple
exercise of the MDM for a dynamically generated resonance, and a full
calculation within a coupled-channel framework remains to be done. In this
respect, the magnetic moment of the $\Lambda(1405)$, which is in the same
lowest energy $J^P = {\frac{1}{2}}^-$ baryon octet as the $S_{11}(1535)$,
has recently been evaluated in Ref.~\cite{Jido02} using unitarized chiral
perturbation theory. In this approach the $\Lambda(1405)$ is dynamically
generated. It would be very worthwhile to also perform such a calculation
for the MDM of the $S_{11}(1535)$ resonance.

\section{Effective Lagrangian model}
\label{sec:Model}%
In this section we develop an effective Lagrangian model for the $\gamma p
\to \gamma \eta p$ reaction in the $S_{11}(1535)$ resonance region, which
will subsequently be used as a tool to investigate the sensitivity to the
$S_{11}(1535)$ MDM.

In the $\gamma p \to \gamma \eta p$ process, a photon ($k$, $\lambda$) hits
a proton target ($p_N$, $s_N$); in the final state, a photon ($k'$,
$\lambda'$), an $\eta$ meson ($q$), and a proton ($p'_N$, $s'_N$) are
observed. Here $k$, $k'$, $p_N$, $p'_N$, and $q$ are the four-momentum for
the respective particles, $\lambda$ and $\lambda'$ denote the photon
helicity, and $s_N$ and $s'_N$ are the proton spin projection.

For convenience, our results for the experimental observables will be
expressed in the center-of-mass (c.m.) frame of the initial $\gamma p$
system. The kinematics of the $\gamma p \to \gamma \eta p$ reaction can be
described by five variables. First, we choose the energies
$E_\gamma^{\mathrm{cm}}$ and $E'^{\mathrm{cm}}_\gamma$ of the initial and
outgoing photon, respectively. The other three variables  are the polar and
azimuthal angles $\theta_\gamma^{\mathrm{cm}}$, $\phi_\gamma^{\mathrm{cm}}$
of the final photon, and $\theta_\eta^{\mathrm{cm}}$ the polar angle of the
eta meson. They are defined by choosing the plane which contains the
initial particles and the final $\eta$ to be the $xz-$plane with $\bm{k}$
pointing in the $z$-direction, and thus $\phi_\eta^{\mathrm{cm}} \equiv 0$.

The unpolarized five-fold differential cross section for the $\gamma p \to
\gamma \eta p$ reaction, which is differential with respect to the outgoing
photon energy and angles as well as the eta meson angles in the c.m.
system, is given by
\begin{eqnarray}
\label{eq:cross3b}%
 \frac{d\sigma}{dE'^{\mathrm{cm}}_\gamma\,
                d\Omega_\gamma^{\mathrm{cm}}\,d\Omega_\eta^{\mathrm{cm}}} \;
&=&\;
 \frac{1}{(2 \pi)^5}\, {\frac{1}{32 \sqrt{s}}}\,
 \frac{E'^{\mathrm{cm}}_\gamma}{E^\mathrm{cm}_\gamma} \,
 \frac{|\,\bm{q}\,|^2}{|\,\bm{q}\,| (E'_p+\omega_q) +
E'^{\mathrm{cm}} \omega_q \cos \theta_{\gamma' \eta} } \, \nonumber\\
&\times&
 \left(\, \frac{1}{4}\sum_{\lambda}\sum_{s_N}\sum_{\lambda'}\sum_{s_N'}
 | \, \varepsilon_\mu(k, \lambda) \, \varepsilon_\nu^*(k',\lambda') \,
 {\mathcal M}^{\nu \mu} \, |^2 \, \right) \, ,
\end{eqnarray}
where $\omega_q$ and $\bm{q}$ are the
energy and momentum of the eta meson, $E'_p$ is the final proton
energy, $\theta_{\gamma' \eta}$ the c.m. angle between the outgoing photon
and the eta, and $\varepsilon_\mu(k,\lambda)$ and
$\varepsilon_\nu^*(k',\lambda')$ are the polarization vectors of the
incoming and outgoing photons, respectively. Furthermore, ${\mathcal
M}^{\nu \mu}$ is a tensor for the $\gamma p \to \gamma \eta p$ process,
which will be calculated next within an effective Lagrangian model.

\begin{figure}
\centering
\includegraphics[width=0.45\columnwidth]{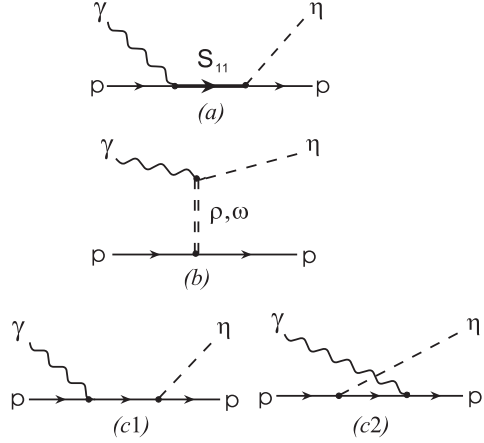}
\caption{\label{fig:gap_etap_diag}Diagrams for the $\gamma p \rightarrow
\eta p$ reaction in the $S_{11}(1535)$ region~: $S_{11}$ resonance
excitation (a), vector meson exchange (b), and Born diagrams (c1,c2). }
\end{figure}

To develop a model for the $\gamma p \to \gamma \eta p$ reaction, we start
from an effective Lagrangian model for the $\eta$ photoproduction reaction
$\gamma p \to \eta p$, whose parameters are determined by a fit to the
experimental data. The fixed parameters are subsequently used to calculate
the $\gamma p \to \gamma \eta p$ process. The dominant contributions for
$\eta$ photoproduction in the $S_{11}(1535)$ region are shown in
Fig.~\ref{fig:gap_etap_diag}. These include, in addition to the resonance
contribution, the nucleon Born terms and the $t$-channel vector meson
exchanges as the nonresonant background. Besides the dominant
$S_{11}(1535)$ nucleon resonance, we also include the $S_{11}(1650)$ since
its contribution in the $S_{11}(1535)$ region is not negligible. Other
nucleon resonances are neglected at this stage since their contributions to
the $\eta$ photoproduction cross sections in the $S_{11}(1535)$ region are
small.

\begin{figure}
\centering
\includegraphics[width=0.65\columnwidth]{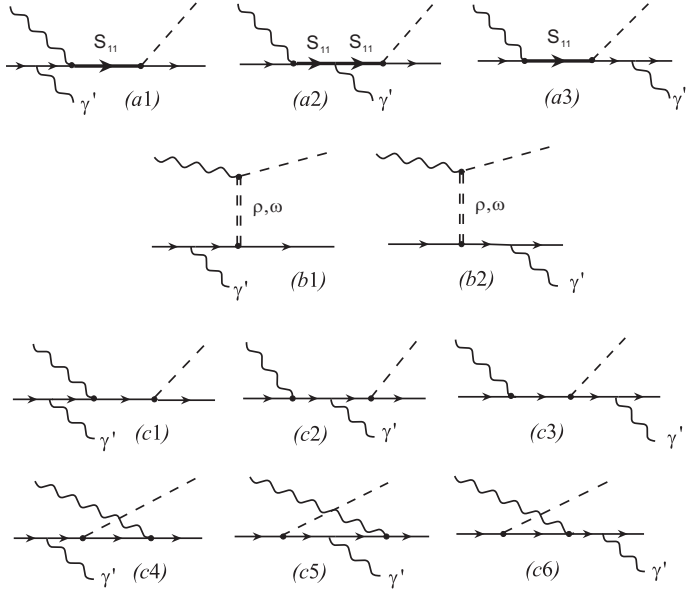}
\caption{\label{fig:gap_etagap_diag}Diagrams considered in the calculation
of the $\gamma p \rightarrow \gamma \eta p$ reaction in the $S_{11}(1535)$
region, obtained by gauge invariant coupling of a photon to the diagrams of
Fig.~\ref{fig:gap_etap_diag}~: $S_{11}$ resonance diagrams (a1-a3),
vector-meson exchange diagrams (b1-b2), and Born diagrams (c1-c6). }
\end{figure}

To extend our description to the $\gamma p \rightarrow \gamma \eta p$
reaction, we couple a photon in a gauge invariant way to all charged
particles in Fig.~\ref{fig:gap_etap_diag}. The resulting diagrams are shown
in Fig.~\ref{fig:gap_etagap_diag}, where the $\gamma$-$S_{11}$-$S_{11}$
vertex in diagram (a2) contains contribution from the magnetic moments of
the $S_{11}$ resonances. Here we assume that the anomalous MDM of the
${S_{11}^+(1650)}$ vanishes, because it has only a very small effect in the
$S_{11}(1535)$ region and its predicted value from the quark model is small
(see Sec.~\ref{sec:QM}). Therefore, in comparison with the $\gamma p \to
\eta p$ process, the only new parameter entering in the description of the
$\gamma p \rightarrow \gamma \eta p$ process is the $S_{11}^+(1535)$
anomalous magnetic moment $\kappa_{S_{11}^+(1535)}$. In the following, we
discuss the $S_{11}$ resonance as well as the background contributions to
the $\gamma p \rightarrow \gamma \eta p$ reaction.

\subsection{$S_{11}(1535)$ resonance propagator}
\begin{figure}
\centering
\includegraphics[width=0.55\columnwidth]{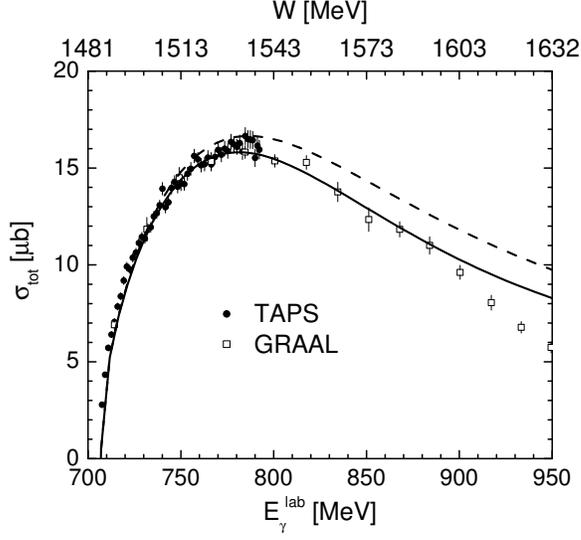}
\caption{\label{fig:getatxs}Total cross section for the $\gamma p
\rightarrow \eta p$. The solid line is the full result, and the dashed line
indicates the contribution from the $S_{11}(1535)$. The data are from
TAPS~\cite{Krusche:1995nv} and GRAAL~\cite{Renard:2000iv}. }
\end{figure}

As the $S_{11}(1535)$ resonance is a spin-1/2 particle, its Feynman
propagator is given by~:
\begin{equation} \label{eq:redprop}
  G_R(p) = \frac{1}{\pdag - M_R} \,,
\end{equation}
where $p$ is the four-momentum and $M_R$ the mass of the $S_{11}$(1535). To
take account of the finite width of the $S_{11}$ resonance, we follow the
procedure of Refs.~\cite{ElAmiri:1992xa,LopezCastro:2000ep} by using a
complex pole description for the resonance excitation. This amounts to the
replacement
\begin{equation} \label{eq:cmplxmass}
M_R \,\longrightarrow\, M_R - \frac{i}{2} \, \Gamma_R \,
\end{equation}
in the propagator of Eq.~(\ref{eq:redprop}). This `complex mass scheme'
guarantees electromagnetic gauge invariance. In contrast, the use of a
Breit-Wigner propagator with an energy-dependent width will violate gauge
invariance when applied to the $S_{11}$ contribution for the $\gamma p
\rightarrow \gamma \eta p$ reaction. By fitting the $\gamma p \rightarrow
\eta p$ data in the $S_{11}(1535)$ region, we can determine the complex
pole of the $S_{11}(1535)$. The fit results are shown in
Fig.~\ref{fig:getatxs}. Limiting our fit to the data near the
$S_{11}(1535)$ peak in the range $E_\gamma^\mathrm{lab}=730-880$~MeV, we
obtain $(M_R,~\Gamma_R)$ = (1490,~154) MeV, which will be used in our
calculation for the $\gamma p \rightarrow \gamma \eta p$ process. In
comparison, the value given by the PDG~\cite{Hagiwara:2002pw} is
(1505,~170) MeV. However, using the PDG value the agreement with the
photoproduction data is less satisfactory.

\subsection{$S_{11}(1535)$ resonance amplitudes}
\label{sec:amp}%
The interaction Lagrangians relevant for the resonance contribution ($R =
S_{11}$) in the $\gamma p \rightarrow \gamma \eta p$ process are
\begin{eqnarray} \label{eq:Lagr}
 \mathcal{L}_{\gamma RR} &=& -e\,\bar\psi_R \left[ \,e_R \gamma_{\mu}A^{\mu}
  - \tfrac{\kappa_R}{2 M_{R}} \sigma_{\mu\nu} \partial^{\nu}
  A^{\mu} \right] \psi_R \,,
\nonumber\\ [0.5ex]
 \mathcal{L}_{\gamma NR} &=& - \tfrac{e}{2(M_N+M_R)} \,\bar\psi_R\,
 \kappa_{NR}\,\gamma_5\,\sigma_{\mu\nu}\,\psi_N F^{\mu\nu}+\mathrm{h.c.}\,,
\nonumber\\ [0.5ex]
 \mathcal{L}_{\eta N R} &=& - \,i \, g_{\eta N R}\,
 \bar{\psi}_N\, \psi_R\, \phi_\eta + \mathrm{h.c.} \,,
\end{eqnarray}
where $g_{\eta N R}$ is the pseudoscalar coupling for the $\eta N R$
vertex, $e_{R}$ and $\kappa_{R}$ the resonance charge and the anomalous
magnetic dipole moment, $\kappa_{NR}$ the $N \rightarrow R$ transition
magnetic coupling, and $F^{\mu\nu} = \partial^\mu A^\nu - \partial^\nu
A^\mu$.

Using these effective Lagrangians, we can derive the amplitudes for the
$S_{11}$ resonance contributions in Fig.~\ref{fig:gap_etap_diag}(a) and
Fig.~\ref{fig:gap_etagap_diag}(a1-a3). The amplitude for the $S_{11}$(1535)
resonance contribution to the $\gamma p \to \eta p$ process, corresponding
to Fig.~\ref{fig:gap_etap_diag}(a), is given by
\begin{eqnarray} \label{eq:getas11}
 & &\varepsilon_\mu(k,\lambda)\,{\mathcal M}^{\mu}_a(\gamma p \to \eta p)\,
\nonumber\\ [0.5ex]
 &=& \,i\,\tfrac{e\,g_{\eta N R}}{M_N + M_R} \,
    \varepsilon_\mu(k, \lambda) \, k_\alpha
    \bar{N}(p'_N, s'_N) \, G_R(p_R) \, \kappa_{pR} \,
           \gamma_5 \, \sigma^{\mu\alpha} \, N(p_N, s_N) \,,
\end{eqnarray}
where $G_R$ is the $S_{11}$ propagator given by Eqs.~(\ref{eq:redprop}) and
(\ref{eq:cmplxmass}).

Starting from Eq.~(\ref{eq:getas11}) to the $\gamma p \to \eta p$ process,
we now construct the corresponding $S_{11}$ contribution to the $\gamma p
\to \gamma \eta p$ process. This is obtained by coupling a photon in a
gauge invariant way to all charged particles in
Fig.~\ref{fig:gap_etap_diag}(a). First we obtain the diagrams with a photon
attached to an external proton [Figs.~\ref{fig:gap_etagap_diag}(a1) and
\ref{fig:gap_etagap_diag}(a3)], described by the usual Dirac and Pauli
currents with $\kappa_p=1.79$ for the proton anomalous MDM. The
contributions of these diagrams to the tensor ${\mathcal M}^{\nu \mu}$ of
Eq.~(\ref{eq:cross3b}) are given by
\begin{eqnarray}
& &{\mathcal M}^{\nu \mu}_{a1} (\gamma p \to \gamma \eta p)
 \nonumber\\ [0.5ex]
&=& \,i\,\tfrac{e^2 g_{\eta N R}}{M_N + M_R}\,k_\alpha\,
  \bar N(p'_N, s'_N) \, G_R(p'_R) \, \kappa_{pR} \,\gamma_5 \,
  \sigma^{\mu\alpha}
 \nonumber\\
& & \times\; S_N(p_N-k') \left[ \gamma^\nu - \kappa_p \,
  i\sigma^{\nu\beta} \tfrac{k'_\beta}{2 M_N} \right] N(p_N, s_N) \,,
  \label{eq:a1}\\ [1ex]
& &{\mathcal M}^{\nu \mu}_{a3} (\gamma p \to \gamma \eta p)
 \nonumber\\ [0.5ex]
&=& \,i\,\tfrac{e^2 g_{\eta N R}}{M_N + M_R}\,k_\alpha\,
   \bar N(p'_N, s'_N) \left[ \gamma^\nu - \kappa_p \,
   i \sigma^{\nu \beta} \tfrac{k'_\beta}{2 M_N} \right]
 \nonumber\\
& & \times\; S_N(p'_N+k')\, G_R(p_R) \, \kappa_{pR} \, \gamma_5 \,
   \sigma^{\mu\alpha} \, N(p_N, s_N) \,, \label{eq:a3}
\end{eqnarray}
where $p'_R$ and $p_R$ correspond to the four-momenta of the $S_{11}$
resonance in Fig.~\ref{fig:gap_etagap_diag}(a1) and
\ref{fig:gap_etagap_diag}(a3). In the soft-photon limit ($k' \to 0$), the
coupling of the photon to the external lines gives the only contribution.
However, at finite energy for the emitted photon, gauge invariance also
requires the diagram with a photon attached to the intermediate $S_{11}$
resonance [Fig.~\ref{fig:gap_etagap_diag}(a2)], which corresponds to the
tensor,
\begin{eqnarray} \label{eq:a2}
& &{\mathcal M}^{\nu \mu}_{a2} (\gamma p \to \gamma \eta p)
 \nonumber\\ [0.5ex]
&=&\,i\,\tfrac{e^2g_{\eta N R}}{M_N + M_R}\,k_\alpha\,
   \bar N(p'_N, s'_N)\,G_R(p'_R)\,\Gamma_{\gamma R R}^{\,\nu}\, G_R(p_R)
 \nonumber\\
& & \times\; \kappa_{pR}\,\gamma_5\,\sigma^{\mu\alpha}\,N(p_N, s_N) \,,
\end{eqnarray}
with $G_R(p_R)$ and $G_R(p'_R)$ the $S_{11}$ propagators
before and after emitting the photon in Fig.~\ref{fig:gap_etagap_diag}(a2),
and the $\gamma R R$ vertex
\begin{equation} \label{eq:gaRR}
  \Gamma_{\gamma R R}^{\,\nu} \,=\, \gamma^\nu\,-\,i\,\kappa_R\,
  \sigma^{\nu \beta}\,\tfrac{k'_\beta}{2 M_R} \,,
\end{equation}
containing the photon coupling to the charge of a spin 1/2 field of a Dirac
particle~\cite{ElAmiri:1992xa} as well as the $S_{11}$ Pauli current
proportional to $\kappa_R$.

The vertex for a point spin-1/2 particle follows from the requirement of
gauge invariance. Indeed, only the sum of the three diagrams (a1,\;a2,\;a3)
in Fig.~\ref{fig:gap_etagap_diag} is gauge-invariant, which is expressed
through the electromagnetic Ward identity relating the $\gamma R R$ vertex
with the $S_{11}$ propagator,
\begin{equation} \label{eq:Wardid}
  \left( p'_R-p_R\right)_\nu\,G(p'_R)\;\Gamma_{\gamma R R}^{\,\nu}\;G(p_R)\,
  = \,G(p_R)\,-\,G(p'_R) \,.
\end{equation}

To include the finite width of the $S_{11}$, the `complex mass scheme' was
advocated in Ref.~\cite{ElAmiri:1992xa}, i.e., the replacement of $M_R$ by
the complex mass of Eq.~(\ref{eq:cmplxmass}). In doing so, it is easily
checked that the Ward identity of Eq.~(\ref{eq:Wardid}) still holds. On the
other hand, when using Breit-Wigner propagators, and replacing $M_R^2$ in
the denominator of Eq.~(\ref{eq:redprop}) by $M_R^2 \,-\, i \, M_R
\,\Gamma_R(W)$, with $\Gamma_R(W)$ an energy dependent width, one
immediately sees that the ${S_{11}}$ propagators before and after the
emission of the photon have different widths (except for the soft-photon
limit $k' \to 0$, where the contribution of diagram
Fig.~\ref{fig:gap_etagap_diag}(a2) vanishes relative to those of
Fig.~\ref{fig:gap_etagap_diag}(a1) and (a3)). Therefore, when using
energy-dependent widths at finite energy of the emitted photon, it is not
possible to maintain the Ward identity of Eq.~(\ref{eq:Wardid}) with the
vertex of Eq.~(\ref{eq:gaRR}).

\subsection{Background contributions}
\label{sec:backgd}%
The effective Lagrangians relevant to the background contribution in the
$\gamma p \rightarrow \gamma \eta p$ process are
\begin{eqnarray} 
\mathcal{L}_{\gamma NN} &=& -e\,\bar\psi_N \left[ \,\hat{e}_N \gamma_{\mu}A^{\mu}
 - \tfrac{\kappa_N}{2 M_{N}} \sigma_{\mu\nu} \partial^{\nu}
 A^{\mu} \right] \psi_N \,,
  \nonumber\\ [0.5ex]
\mathcal{L}_{\eta N N} &=& - i\, g_{\eta N N}\, \bar{\psi}_N\, \gamma_5\,
\psi_N\, \phi_\eta \,,
  \nonumber\\ [0.5ex]
{\mathcal{L}}_{\gamma \eta V} & = & \tfrac{e \lambda_{\gamma \eta
V}}{m_{\eta}}\, \varepsilon_{\mu \nu \rho \sigma}\,(\partial^{\mu}
A^{\nu})\,\phi_{\eta}\, (\partial^{\rho} V^{\sigma}) \,,
  \nonumber\\ [0.5ex]
{\mathcal{L}}_{VNN} & = & g_{VNN} {\bar{\psi}_N} \left( \gamma_{\mu}
V^{\mu} + \tfrac{\kappa_V}{2 m_N} \sigma_{\mu \nu}
\partial^{\nu} V^{\mu} \right) \, \psi_N \,,
  \label{eq:Lag}
\end{eqnarray}
where $V$ denotes $\rho$ and $\omega$ vector meson fields, and the
pseudoscalar coupling $g_{\eta NN}$ is used in our calculation. With these
effective Lagrangians, it is straightforward to derive the amplitudes
corresponding to the diagrams of Fig.~\ref{fig:gap_etap_diag}(b) and
(c1,c2) as well as Fig.~\ref{fig:gap_etagap_diag}(b1,b2) and (c1-c6).

For the $\eta NN$ coupling in Eq.~(\ref{eq:Lag}), we determine its value by
fitting the photoproduction data and obtain $g_{\eta NN}^2 / 4 \pi =0.05$.
For the vector meson couplings, we use the values from
Ref.~\cite{Chiang:2001as}. There are uncertainties associated with the
$g_{\eta NN}$ and vector meson couplings, and it could affect the correct
extraction of the $S_{11}(1535)$ MDM. However, the background
contributions, which contain the Born terms and vector meson exchanges, do
not exceed $1\%$ of the $\gamma p \to \eta p$ and $\gamma p \to \gamma \eta
p$ cross sections in the $S_{11}(1535)$ energy region considered here.
Therefore, we believe that these uncertainties will not significantly
affect the extraction of the $S_{11}(1535)$ MDM.

\section{Soft photon limit}
\label{sec:soft}%
In the soft photon limit ($k'\rightarrow 0$), gauge invariance provides a
model-in\-de\-pen\-dent relation between the cross sections for the $\gamma
p \rightarrow \gamma \eta p$ and $\gamma p \rightarrow \eta p$ processes.
In this section, we formulate this low energy theorem. The expression we
derive applies to the production of any neutral meson $M$, and we therefore
refer more generally to the $\gamma p \rightarrow \gamma M p$ reaction in
this section. In particular, it also applies to the $\gamma p \rightarrow
\gamma \pi^0 p$ reaction as studied in Ref.~\cite{Drechsel:2001qu}.

All variables in this section are expressed in the c.m.~frame of
the initial $\gamma p$ system. In the soft photon limit, the momentum and
energy of the outgoing photon vanishes, and the $\gamma p \rightarrow
\gamma M p$ reaction is dominated by the bremsstrahlung process from the
initial and final protons.
In this limit, when integrating the five-fold
differential cross sections for the $\gamma p \rightarrow \gamma M p$ process
over the outgoing photon angles, we obtain
(the details are given in Appendix~\ref{app:soft})
\begin{equation} \label{eq:soft}
  \frac{d\sigma}{dE'^{\mathrm{cm}}_\gamma d\Omega^{\mathrm{cm}}_M}
  \longrightarrow
  \frac{1}{E'^{\mathrm{cm}}_\gamma} \frac{e^2}{2 \pi^2} \,W(v)
  \left(\frac{d\sigma}{d\Omega_M}\right)^{\!\mathrm{cm}}
  \quad \mathrm{as} \ k'\rightarrow 0 \,.
\end{equation}
Here $(d\sigma/d\Omega_M)^{\!\mathrm{cm}}$ is the differential cross
section for the $\gamma p \rightarrow M p$ process, and we define an
angular weight-function
\begin{equation} \label{eq:wv}
  W(v)
  \equiv
  -1+\left(\frac{v^2+1}{2v}\right) \ln\left(\frac{v+1}{v-1}\right) \,,
\end{equation}
where $v \equiv \sqrt{1+\frac{4M_N^2}{-t}}$ with $t=(k-q)^2$ and $M_N$ the
nucleon mass.

If we integrate the cross section of Eq.~(\ref{eq:soft}) also over the
meson angles, the result is
\begin{eqnarray} \label{eq:softint}
  \frac{d\sigma}{dE'^{\mathrm{cm}}_\gamma}
  &\equiv&
  \int d\Omega^{\mathrm{cm}}_M
  \frac{d\sigma}{dE'^{\mathrm{cm}}_\gamma d\Omega^{\mathrm{cm}}_M} \,,
  \\ \nonumber
  &\longrightarrow&
  \frac{1}{E'^{\mathrm{cm}}_\gamma} \cdot \bar{\sigma}_M
  \quad \mathrm{as} \ k'\rightarrow 0 \,,
\end{eqnarray}
with a ``weight-averaged'' total cross section,
\begin{equation} \label{eq:getaspl}
  \bar{\sigma}_M
  \equiv
  \frac{e^2}{2 \pi^2} \int d\Omega^{\mathrm{cm}}_M\, W(v)
  \left(\frac{d\sigma}{d\Omega_M}\right)^{\!\mathrm{cm}}
  \,.
\end{equation}

\begin{figure}
\centering
\vspace{-1cm}
\includegraphics[width=0.55\columnwidth]{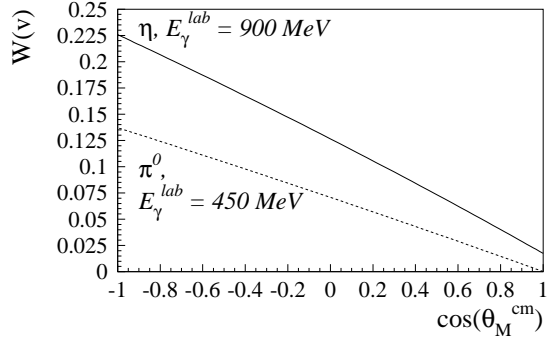}
\caption{\label{fig:bremspieta}Angular weight-function $W(v)$ of
Eq.~(\ref{eq:wv}) in the soft-photon formula for the $\gamma
p \rightarrow \gamma \eta p$ (solid curve) and $\gamma p
\rightarrow \gamma \pi^0 p$ (dashed curve) processes,
as function of the meson angle $\theta_M^{cm}$. }
\end{figure}

We show the angular weight-function $W(v)$ in Fig.~\ref{fig:bremspieta} for
both the $\gamma p \rightarrow \gamma \eta p$ and the $\gamma p \rightarrow
\gamma \pi^0 p$ processes. If the meson is emitted in the forward direction
with respect to the initial photon direction, there appears a destructive
interference between the bremsstrahlung from the initial and final
nucleons, whereas for a meson emitted in the backward direction both
bremsstrahlung sub-processes add constructively.

The low energy theorem of Eq.~(\ref{eq:softint}) provides a check for both
theoretical model calculations and experimental measurements, because the
ratio
\begin{eqnarray}
\label{eq:rratio}
R \,\equiv \,
E'^{\mathrm{cm}}_\gamma \cdot (d\sigma/dE'^{\mathrm{cm}}_\gamma)
/ \bar{\sigma}_M ,
\end{eqnarray}
with $d\sigma/dE'^{\mathrm{cm}}_\gamma$
for $\gamma p \rightarrow \gamma M p$ and $\bar{\sigma}_M$
for $\gamma p \rightarrow M p$, approaches unity when $k' \rightarrow 0$.

\begin{figure}
\centering
\includegraphics[width=0.55\columnwidth]{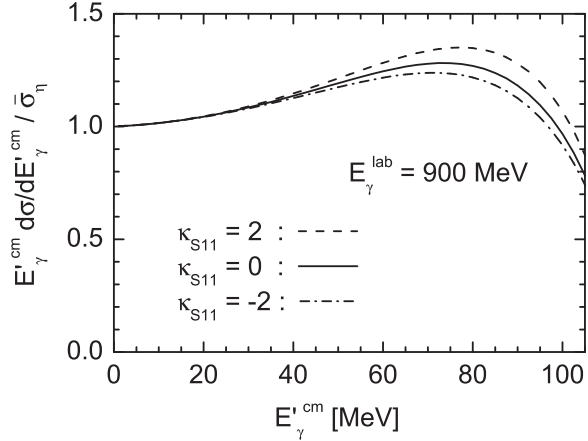}
\caption{\label{fig:splcheck}Outgoing photon energy dependence of the
ratio $R$, defined in Eq.~(\ref{eq:rratio}),
for the $\gamma p \rightarrow \gamma \eta p$ reaction at
$E_\gamma^{\mathrm{lab}}=900$~MeV. }
\end{figure}

In Fig.~\ref{fig:splcheck}, we show the ratio $R$ constructed from our
theoretical calculations of the $\gamma p \rightarrow \gamma \eta p$ and
$\gamma p \rightarrow \eta p$ reactions. We indeed observe that $R$
approaches unity when $k' \rightarrow 0$. For $k' \simeq 20$~MeV, this
ratio is still within 5\% of the soft photon limit. Upon increasing the
outgoing photon energy further, one starts to see clear deviations (of the
order of 30\%) from the soft photon limit. These deviations allow one to
extract new resonance information from the $\gamma p \rightarrow \gamma
\eta p$ reaction, which will be studied in the next section.

\section{Results for $\bm{\gamma\,p\to\gamma\,\eta\,p}$ observables}
\label{sec:results}%
In this section, we use our effective Lagrangian model as a tool to study
the feasibility of an experiment to measure the $\gamma p \to \gamma \eta
p$ reaction. In particular, we present our results for the $\gamma p \to
\gamma \eta p$ cross sections and photon asymmetries for different values
of $\kappa_{S_{11}(1535)}$ in the range of the estimates made in
Section~\ref{sec:modelmdm}, i.e., between $-2$ to $+2$ (in units of
$\mu_N$), to illustrate the sensitivity with respect to the $S_{11}(1535)$
MDM.

\begin{figure}
\centering
\includegraphics[width=0.55\columnwidth]{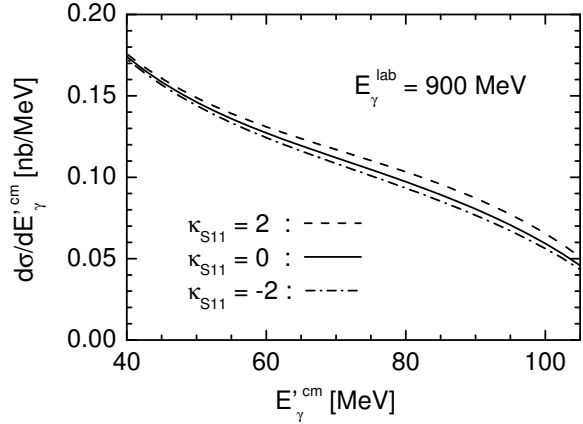}
\caption{\label{fig:xsvsE900}Outgoing photon energy dependence of the cross
section $d\sigma/dE'^{\mathrm{cm}}_\gamma$ (integrated over photon and eta
angles) for the $\gamma p \rightarrow \gamma \eta p$ reaction at
$E_\gamma^{\mathrm{lab}}=900$~MeV. }
\end{figure}

In Fig.~\ref{fig:xsvsE900}, we show the outgoing photon energy dependence
of the cross section $d\sigma/dE'^{\mathrm{cm}}_\gamma$ for the $\gamma p
\rightarrow \gamma \eta p$ reaction integrated over the photon and eta
angles for fixed $E_\gamma^{\mathrm{lab}}=900$~MeV. The cross section at
low energies $E'^{\mathrm{cm}}_\gamma$ exhibits the characteristic
bremsstrahlung behavior, but displays clear deviations from such a behavior
when going to higher outgoing photon energies (see also
Fig.~\ref{fig:splcheck}). To give an idea about the feasibility to measure
the $\gamma p \rightarrow \gamma \eta p$ cross section, it is worthwhile to
make the comparison with the $\gamma p \to \gamma \pi^0 p$ reaction in the
$\Delta(1232)$ region, which has already been measured \cite{Kot99}. At
zero outgoing photon energy, the cross section for radiative eta production
can be estimated from the $\gamma p \rightarrow \eta p$ cross section. On
the other hand, the maximum value of the $\gamma p \rightarrow \eta p$
cross section at the $S_{11}(1535)$ position is about 5~\% of the $\gamma p
\rightarrow \pi^0 p$ cross section at the $\Delta(1232)$ position. However,
at an outgoing photon energy of about 100 MeV, which is the region where it
is sensitive to the MDM of the resonance, our estimate for the $\gamma p
\to \gamma \eta p$ reaction at $E_\gamma^{\mathrm{lab}}=900$~MeV yields
count rates that are about 10--15~\% of the measured value of
Ref.~\cite{Kot99} in case of the $\gamma p \to \gamma \pi^0 p$ reaction at
$E_\gamma^{\mathrm{lab}}=450$~MeV, which turns out to be much more
favorable than the 5~\% found in the soft photon limit.

\begin{figure}
\centering
\includegraphics[width=0.55\columnwidth]{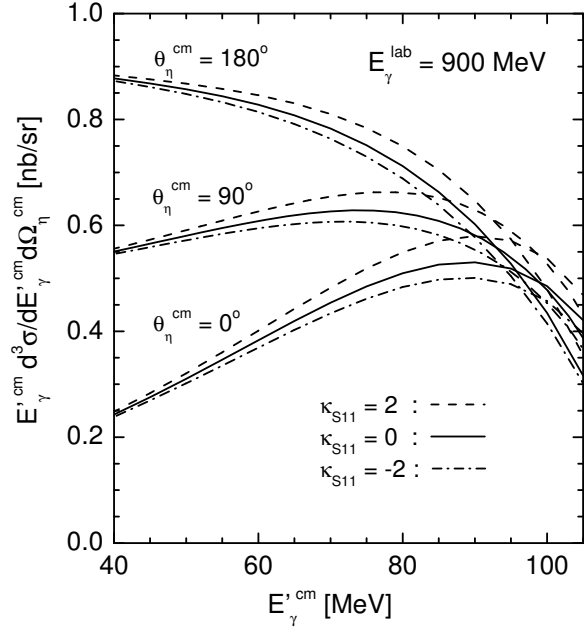}
\caption{\label{fig:xs3f900}Outgoing photon energy dependence of the
$E'^{\mathrm{cm}} \cdot d^3\sigma / dE'^{\mathrm{cm}}_\gamma
d\Omega^{\mathrm{cm}}_\eta$ for the $\gamma p \rightarrow \gamma \eta p$
reaction at $E_\gamma^{\mathrm{lab}}=900$~MeV. }
\end{figure}

One can increase the sensitivity of the $\gamma p \rightarrow \gamma \eta
p$ reaction to the $S_{11}(1535)$ MDM by measuring differential cross
sections in specific kinematical regions. In Fig.~\ref{fig:xs3f900}, we
show the cross section, in the c.m. frame of the initial $\gamma p$ system,
for different values of the $\eta$ angle upon integration over all photon
directions at $E_\gamma^{\mathrm{lab}}=900$~MeV. At the lower energies, one
sees a strong forward-backward asymmetry, which was discussed before in
Fig.~\ref{fig:bremspieta}, yielding substantially larger cross sections for
backward $\eta$ c.m. angles. However, at $E'^{\mathrm{cm}}_\gamma \approx
100$~MeV, the cross sections become comparable for all angles and exhibit a
20\% change when $\kappa_{S_{11}(1535)}$ is varied between $-2$ to $+2$.

\begin{figure}
\centering
\includegraphics[width=\columnwidth]{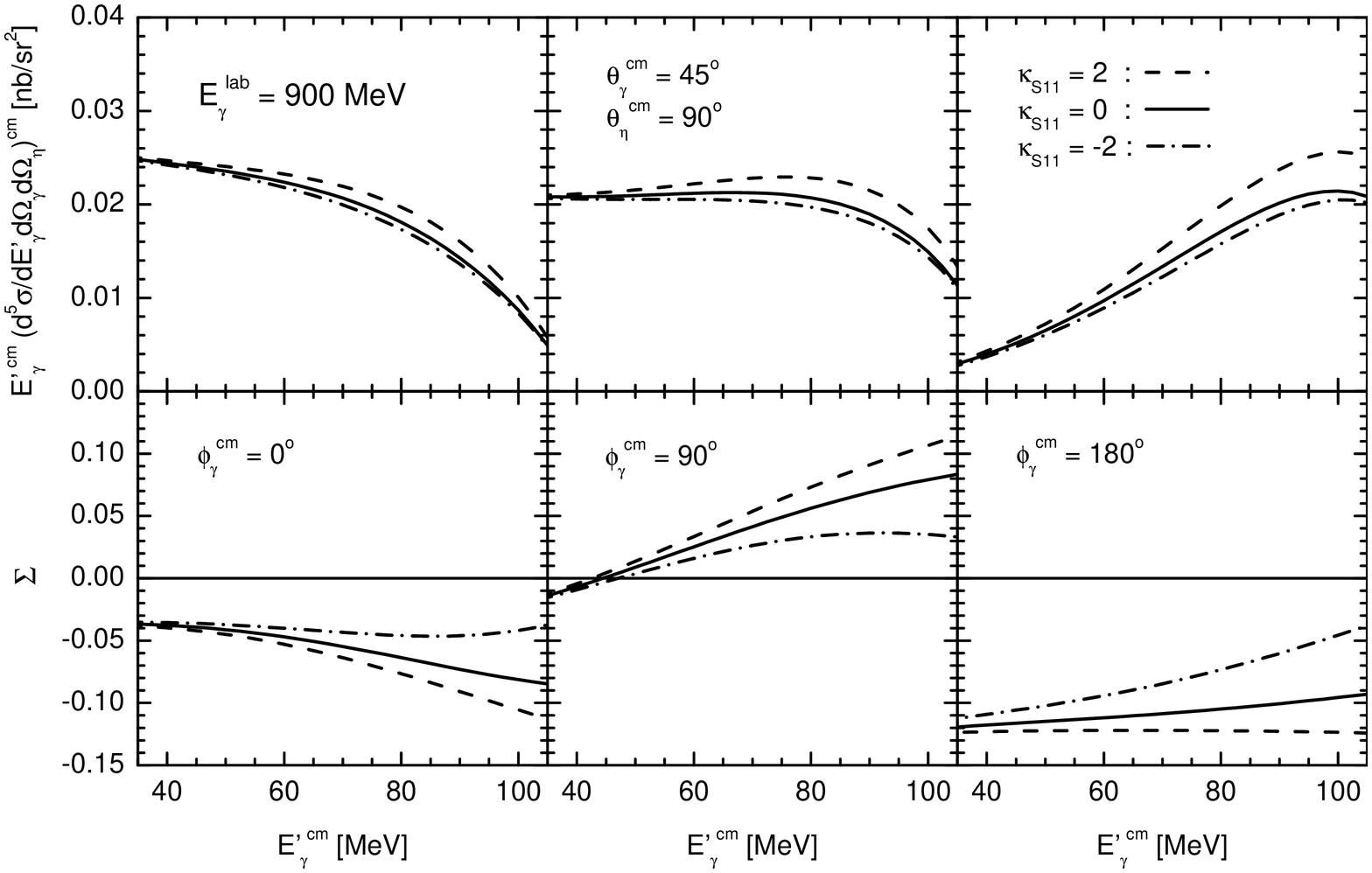}
\caption{\label{fig:5f900}Outgoing photon energy dependence of
$E'^{\mathrm{cm}} \cdot d^5\sigma/dE'^{\mathrm{cm}}_\gamma
d\Omega^{\mathrm{cm}}_\gamma d\Omega_\eta^{\mathrm{cm}}$ and photon
asymmetry $\Sigma$ of the $\gamma p \rightarrow \gamma \eta p$ reaction for
fixed polar angles $\theta_\eta^{\mathrm{cm}}=90^\circ$ and
$\theta_\gamma^{\mathrm{cm}}=45^\circ$, and varying azimuthal angle
$\phi_\gamma^{\mathrm{cm}}=0^\circ,\;90^\circ,\;180^\circ$ (from left to
right) at $E_\gamma^{\mathrm{lab}}=900$~MeV. }
\end{figure}

One can further increase the sensitivity by measuring cross sections at
particular angular regions of the emitted $\eta$ and $\gamma$. As an
example, we show in Fig.~\ref{fig:5f900} the five-fold c.m. differential
cross section $d^5\sigma/dE'^{\mathrm{cm}}_\gamma
d\Omega_\gamma^{\mathrm{cm}} d\Omega_\eta^{\mathrm{cm}}$ and the photon
asymmetry $\Sigma$ at $E_\gamma^{\mathrm{lab}}=900$~MeV for the polar
angles $\theta_\eta^{\mathrm{cm}}=90^\circ$ and
$\theta_\gamma^{\mathrm{cm}}=45^\circ$, and for three values of the
azimuthal angle $\phi_\gamma^{\mathrm{cm}}=0^\circ,\;90^\circ,\ {\rm{and}}\
180^\circ$ (note that the $\eta$ is emitted in the plane, i.e.,
$\phi_\eta^{\mathrm{cm}}\equiv0^\circ$). We notice that the photon
asymmetries reach values of about 10\% around $E'^{\mathrm{cm}}_\gamma =
100$~MeV, and show a strong dependence on the out-of-plane angle of the
emitted photon. One also finds that the sensitivity of the photon asymmetry
to the value of $\kappa_{S_{11}(1535)}$ is quite sizeable.

\section{Summary and conclusions}
\label{sec:sum}%
In this paper we investigated the $\gamma p \to \gamma \eta p$ reaction in
the $S_{11}$(1535) resonance region as a tool to access the $S_{11}$(1535)
magnetic dipole moment (MDM). We performed calculations of the $\gamma p
\to \gamma \eta p$ process using an effective Lagrangian approach, which
contains the $S_{11}(1535)$ and $S_{11}(1650)$ resonant mechanism as well
as a background of non-resonant contributions. As a starting point, the
effective Lagrangian formalism is first applied to the $\gamma p \to \eta
p$ reaction. When extending the calculation to the $\gamma p \to \gamma
\eta p$ process, a photon is coupled to all charged particles in the
$\gamma p \to \eta p$ process, and the coupling is constrained by gauge
invariance. In particular, to take account of the finite width of the
$S_{11}$ resonance, a complex mass scheme rather than a Breit-Wigner
propagator with energy-dependent widths, has been used to maintain the Ward
identity.

We presented our results for the $\gamma p \to \gamma \eta p$ process using
both unpolarized and polarized photon beams. In particular, we focused on
the sensitivity of cross sections and photon asymmetries to the
$S_{11}$(1535) anomalous MDM. We derived a low--energy theorem relating the
cross sections for the $\gamma p \to \eta p$ and $\gamma p \to \gamma \eta
p$ processes in the soft-photon limit, which provides a useful check for
both theoretical calculations and experimental measurements.

We first showed the unpolarized cross section integrated over the full
photon and $\eta$ angles. At low energies this cross section is dominated
by the bremsstrahlung contribution, which is consistent with our check in
the soft-photon limit. The sensitivity of the unpolarized integrated cross
section to $\kappa_{S_{11}(1535)}$ is found to be small, but can be
increased by measuring differential cross sections at particular angular
ranges. For the differential cross section
$d^3\sigma/dE'^{\mathrm{cm}}_\gamma d\Omega_\eta^{\mathrm{cm}}$ integrated
over the photon angles (but not the $\eta$ angles), and the five-fold
differential cross section $d^5\sigma/dE'^{\mathrm{cm}}_\gamma
d\Omega^{\mathrm{cm}}_\gamma d\Omega_\eta^{\mathrm{cm}}$, we found that the
change when varying $\kappa_{S_{11}(1535)}$ between $-2$ to $+2$ is at the
20\%--25\% level. One can further increase the sensitivity by measuring
cross sections using a linearly polarized photon beam. Indeed, our results
clearly show that the sensitivity of the photon asymmetry to
$\kappa_{S_{11}(1535)}$ is quite sizeable.

One inspiring motivation to study the MDMs of the $S_{11}(1535)$ resonance
is to clarify the nature of this resonance---whether it is a traditional
$qqq$ state, or dynamically generated from meson-baryon interactions. In
the framework of the SU(6) quark model, the $S_{11}(1535)$ and
$S_{11}(1650)$ resonances are configuration mixtures of two SU(6) states
with excited orbital wavefunctions. Calculating both the quark spin and
orbital angular momentum contribution for the MDM, as well as the cross
terms due to the configuration mixing, we obtained the MDM values for the
$S_{11}(1535)$ resonance. In comparison, we also presented a simple MDM
calculation, where the $S_{11}$(1535) is dynamically generated and
described as a ($K\Sigma$) molecule. Of course, further predictions of
$S_{11}(1535)$ MDMs from other baryon resonance calculations would be most
welcome, since through this reaction we can investigate the MDMs of
orbitally excited nucleon resonances for the first time by an experiment.
Although challenging, such investigations seem now feasible at ELSA and
MAMI. These planned experiments will provide an interesting test for every
baryon structure calculation, and give insight into the nature of the
parity partner of the nucleon.

\begin{ack}
The authors would like to thank M.M.~Giannini and E.~Santopinto for their
communication of an independent check of the quark model calculation of
Section~\ref{sec:QM}. W.-T.~C. is grateful to the Universit\"at Mainz for
the hospitality extended to him during his visits, and M.Vdh is grateful to
the National Taiwan University for the hospitality during his visits. This
work was supported in parts by the National Science Council of ROC under
grant No.~NSC90-2112-M002-032, by the Deutsche Forschungsgemeinschaft (SFB
443), and by a joint project NSC/DFG TAI-113/10/0.
\end{ack}

\appendix

\section{Quark model calculation}
\label{append:qm}%
To perform a calculation in a constituent quark model with harmonic wave
functions, it is convenient to work with Jacobi coordinates, which are
defined in terms of the three quark positions $\bm{r_i}$,
\begin{eqnarray} 
 \bm{R} &=& \tfrac{1}{\;3\;}\,(\bm{r_1}+\bm{r_2}+\bm{r_3}) \,, \\
 \bm{\rho} &=& \tfrac{1}{\sqrt{2}}\,(\bm{r_1}-\bm{r_2}) \,, \nonumber \\
 \bm{\lambda} &=& \tfrac{1}{\sqrt{6}}\,(\bm{r_1}+\bm{r_2}-2\bm{r_3})
                  \,.\nonumber
\end{eqnarray}
Their inversion relations are
\begin{eqnarray} 
 \bm{r_1} &=& \bm{R} + \tfrac{1}{\sqrt{2}}\,\bm{\rho}
                     + \tfrac{1}{\sqrt{6}}\,\bm{\lambda} \,, \\
 \bm{r_2} &=& \bm{R} - \tfrac{1}{\sqrt{2}}\,\bm{\rho}
                     + \tfrac{1}{\sqrt{6}}\,\bm{\lambda} \,, \nonumber \\
 \bm{r_3} &=& \bm{R} - \sqrt{\tfrac{2}{3}}\,\bm{\lambda} \,. \nonumber
\end{eqnarray}
We usually calculate in the center-of-mass system of the three quarks in a
baryon, i.e., $\bm{R} \equiv 0$.

In a constituent quark model, the MDM of a baryon consists of contributions from both
quark spin and orbital angular momentum,
\begin{eqnarray} \label{eq:mdm}
  \bm{\mu} &=& \bm{\mu^S} + \bm{\mu^L} \,,
\\ [2ex]
  \bm{\mu^S}
  &=& \sum_i \bm{\mu^s}_i
   =  \sum_i \frac{Q_i}{m_i}\, \bm{s}_i \,,
\\ 
  \bm{\mu^L}
  &=& \sum_i \bm{\mu^l}_i
   =  \sum_i \frac{Q_i}{2m_i}\, \bm{l}_i \,, \\ \nonumber
  &=& \sum_i \frac{Q_i}{2m_i}\, \bm{r}_i \times {\bm{p}}_i \,,  \\ \nonumber
  &=& \frac{Q_1+Q_2}{4m}\,\bm{l_\rho}
    + \Bigl( \frac{Q_1+Q_2}{12m}+\frac{Q_3}{3m'} \Bigr)\,\bm{l_\lambda}
    + \frac{Q_1-Q_2}{4\sqrt{3}m}\,\bigl(
      \bm{\rho}\times{\bm{p_\lambda}}+\bm{\lambda}\times{\bm{p_\rho}}
      \bigr) \,,
\end{eqnarray}
where $Q_i$ denotes the charge of the $i$th quark, the index $i$ sums over
three quarks, and $\bm{l_\rho}=\bm{\rho}\times{\bm{p_\rho}}$,
$\bm{l_\lambda}=\bm{\lambda}\times{\bm{p_\lambda}}$.

The calculation for the quark spin part of baryon MDMs is straightforward.
But the calculation for the orbital angular momentum part is much less
familiar, since it vanishes for ground state octet and decuplet baryons,
which have no orbital excitation. To calculate the matrix elements of
$\mu^L_z$ with wavefunctions of Eq.~(\ref{eq:N24}), we need to operate on
spatial wavefunctions $\psi^{\rho,\lambda}_{1m}$ with $z$-components of
$\bm{l_\lambda}$,  $\bm{l_\rho}$, $\bm{\rho}\times{\bm{p_\lambda}}$, and
$\bm{\lambda}\times{\bm{p_\rho}}$. We list in the following the properties
where $\mu^L_z$ acts on spatial wavefunctions $\psi_{1m}$,
\begin{equation} 
  \left. \begin{array}{rclcrcl}
  l_{\rho z}\, \psi^\rho_{1m} &=& m\, \psi^\rho_{1m} \,,
  &\quad& l_{\rho z}\, \psi^\lambda_{1m} &=& 0 \,, \\ [1ex]
  l_{\lambda z}\, \psi^\lambda_{1m} &=& m\, \psi^\lambda_{1m} \,,
  &\quad& l_{\lambda z}\, \psi^\rho_{1m} &=& 0 \,, \\ [1ex]
  (\bm{\rho} \times \bm{p_\lambda})_z\, \psi^\lambda_{1m} &=& m\, \psi^\rho_{1m} \,,
  &\quad& (\bm{\rho} \times \bm{p_\lambda})_z\, \psi^\rho_{1m} &=& 0 \,, \\ [1ex]
  (\bm{\lambda} \times \bm{p_\rho})_z\, \psi^\rho_{1m} &=& m\, \psi^\lambda_{1m} \,,
  &\quad& (\bm{\lambda} \times \bm{p_\rho})_z\, \psi^\lambda_{1m} &=& 0 \,.
          \end{array}
  \right.
\end{equation}
In the case of a symmetric $qqq$ state $|\,(qqq)_s\,\rangle$, the calculation of
$\bm{\mu^L}$ can be simplified as
\begin{equation} \label{eq:muL}
  \langle\,(qqq)_s\,|\,\bm{\mu^L}\,|\,(qqq)_s\,\rangle
 =\langle\,(qqq)_s\,|\,\frac{Q_3}{m_3}\,\bm{l_\lambda}\,|\,(qqq)_s\,\rangle \,.
\end{equation}

In the Tables~\ref{tbl:musp}-\ref{tbl:mul0}, we list the matrix elements
for quark spin and orbital angular momentum contributions of the MDMs for
the states $|\, N^2 P_{1/2} \,\rangle$ and $|\, N^4 P_{1/2} \,\rangle$
with $J_z=+1/2$.

\begin{table}[ht!]
\renewcommand{\tabcolsep}{10mm}
\caption{\label{tbl:musp}Matrix elements for $\langle\, N^{2S+1}P_{1/2}{}^+
\,|\, \mu^S_z \,|\, N^{2S+1}P_{1/2}{}^+ \,\rangle$}
\begin{tabular}{ | c | c c | }
\hline
 $\mu^S_z$ & $|\, N^2P_{1/2}{}^+ \,\rangle$  &
             $|\, N^4P_{1/2}{}^+ \,\rangle$  \\
\hline
 $\langle\, N^2P_{1/2}{}^+ \,|$  &
 $-\frac{1}{9}\,(2\mu_u + \mu_d)$  &
 $\frac{4}{9}\,(\mu_u - \mu_d)$   \\
 $\langle\, N^4P_{1/2}{}^+ \,|$ &
 $\frac{4}{9}\,(\mu_u - \mu_d)$  &
 $\frac{5}{9}\,(2\mu_u + \mu_d)$  \\
\hline
\end{tabular}
\end{table}

\begin{table}[ht!]
\renewcommand{\tabcolsep}{10mm}
\caption{\label{tbl:mus0}Matrix elements for $\langle\, N^{2S+1}P_{1/2}{}^0
\,|\, \mu^S_z \,|\, N^{2S+1}P_{1/2}{}^0 \,\rangle$}
\begin{tabular}{ | c | c c | }
\hline
 $\mu^S_z$ & $|\, N^2P_{1/2}{}^0 \,\rangle$  &
             $|\, N^4P_{1/2}{}^0 \,\rangle$  \\
\hline
 $\langle\, N^2P_{1/2}{}^+ \,|$  &
 $-\frac{1}{9}\,(2\mu_d + \mu_u)$  &
 $\frac{4}{9}\,(\mu_d - \mu_u)$   \\
 $\langle\, N^4P_{1/2}{}^+ \,|$ &
 $\frac{4}{9}\,(\mu_d - \mu_u)$  &
 $\frac{5}{9}\,(2\mu_d + \mu_u)$  \\
\hline
\end{tabular}
\end{table}
\begin{table}[ht!]
\renewcommand{\tabcolsep}{10mm}
\caption{\label{tbl:mulp}Matrix elements for $\langle\, N^{2S+1}P_{1/2}{}^+
\,|\, \mu^L_z \,|\, N^{2S+1}P_{1/2}{}^+ \,\rangle$}
\begin{tabular}{ | c | c c | }
\hline
 $\mu^L_z$ & $|\, N^2P_{1/2}{}^+ \,\rangle$  &
             $|\, N^4P_{1/2}{}^+ \,\rangle$  \\
\hline
 $\langle\, N^2P_{1/2}{}^+ \,|$  &
 $\tfrac{1}{9}\bigl(2\tfrac{Q_u}{m_u}+\tfrac{Q_d}{m_d}\bigr)$  &  0
 \\
 $\langle\, N^4P_{1/2}{}^+ \,|$ &
 0  &  $-\tfrac{1}{18}\bigl(\tfrac{Q_u}{m_u}+2\tfrac{Q_d}{m_d}\bigr)$
 \\
\hline
\end{tabular}
\end{table}
\begin{table}[ht!]
\renewcommand{\tabcolsep}{10mm}
\caption{\label{tbl:mul0}Matrix elements for $\langle\, N^{2S+1}P_{1/2}{}^0
\,|\, \mu^L_z \,|\, N^{2S+1}P_{1/2}{}^0 \,\rangle$}
\begin{tabular}{ | c | c c | }
\hline
 $\mu^L_z$ & $|\, N^2P_{1/2}{}^0 \,\rangle$  &
             $|\, N^4P_{1/2}{}^0 \,\rangle$  \\
\hline
 $\langle\, N^2P_{1/2}{}^0 \,|$  &
 $\tfrac{1}{9}\bigl(2\tfrac{Q_d}{m_d}+\tfrac{Q_u}{m_u}\bigr)$  &  0
 \\
 $\langle\, N^4P_{1/2}{}^0 \,|$  &
 0  &  $-\tfrac{1}{18}\bigl(\tfrac{Q_d}{m_d}+2\tfrac{Q_u}{m_u}\bigr)$
 \\
\hline
\end{tabular}
\end{table}

\section{Derivation of the low energy theorem relating the
$\bm{\gamma p \to \gamma M p}$ and $\bm{\gamma p \to M p}$ processes}
\label{app:soft}%
In the limit $k' \to 0$, the $\gamma p \rightarrow \gamma M p$ reaction is
exactly described by the bremsstrahlung process from the initial and final
protons. This yields the five-fold differential c.m. cross section
\begin{eqnarray} \label{eq:softa1}
  &&\frac{d\sigma}{dE'^{\mathrm{cm}}_\gamma d\Omega^{\mathrm{cm}}_M
                 d\Omega^{\mathrm{cm}}_\gamma} \\\nonumber
  &\longrightarrow&
  \left(\frac{d\sigma}{d\Omega_M}\right)^{\!\mathrm{cm}} \,\cdot \,
  \frac{e^2}{16 \pi^3} \, E'^{\mathrm{cm}}_\gamma \,\cdot \,
  \sum_{\lambda_\gamma} \left| \,
  \frac{p'_N \cdot \varepsilon(k', \lambda_\gamma)}{p'_N \cdot k'}
  \;-\;
  \frac{p_N \cdot \varepsilon(k', \lambda_\gamma)}{p_N \cdot k'}
  \, \right|^2 ,
\end{eqnarray}
as $ k'\rightarrow 0$. In Eq.~(\ref{eq:softa1}), $(d\sigma /
d\Omega_M)^{\mathrm{cm}}$ is the c.m. cross section for the $\gamma p
\rightarrow M p$ process, $\lambda_\gamma = \pm 1$ is the soft photon
polarization, and $\varepsilon$ its polarization vector. We calculate the
RHS of Eq.~(\ref{eq:softa1}), by performing the sum over the photon
polarizations, and integrating over the photon angles. This gives the
result~:
\begin{equation}
\label{eq:softa2}
  \frac{d\sigma}{dE'^{\mathrm{cm}}_\gamma d\Omega^{\mathrm{cm}}_M }
 \longrightarrow
\left(\frac{d\sigma}{d\Omega_M}\right)^{\!\mathrm{cm}} \,\cdot \,
  \frac{e^2}{16 \pi^3} \, E'^{\mathrm{cm}}_\gamma \,\cdot
   \; I \, ,
\end{equation}
where we introduced the photon angular integral $I$ as~:
\begin{equation} \label{eq:softa3}
  I \,\equiv \, \int d \Omega^{\mathrm{cm}}_\gamma \,
  \left[ \, \frac{2 \, p_N \cdot p'_N}{(p_N \cdot k') \, (p'_N \cdot k')}
  \,-\, \frac{M^2}{(p_N \cdot k')^2} \,-\, \frac{M^2}{(p'_N \cdot k')^2} \,
  \right] .
\end{equation}
In Eq.~(\ref{eq:softa3}), the second and third terms arise from the
contribution of brems\-strah\-lung from the same proton, whereas the first
term stems from the interference between the bremsstrahlung from the
initial and final protons.

We next work out the photon angular integral of Eq.~(\ref{eq:softa3}). It
turns out to be useful to introduce the initial and final nucleon
velocities $\bm{\beta_N} \equiv \bm{p_N} / E_N$ and $\bm{\beta'_N} \equiv
\bm{p'_N} / E'_N$, and a Feynman parametrization in the first term of
Eq.~(\ref{eq:softa3}), which brings the two propagator factors to the same
denominator. This leads to the expression~:
\begin{eqnarray} \label{eq:softa4}
  I \,=&& \, \frac{2 \pi}{(E'^{\mathrm{cm}}_\gamma)^2 }
  \cdot \left\{ (1 - \bm{\beta_N} \cdot \bm{\beta'_N}) \, \int_{-1}^{+1}
  dy \, \int_{-1}^{+1} dx \, \frac{1}{(1 - \beta_y \, x)^2} \right.
  \\ \nonumber
&& \ \left.
  \,-\, (1 - \beta_N^2) \int_{-1}^{+1} dx \, \frac{1}{(1 - \beta_N \, x)^2}
  \,-\, (1 - \beta'^2_N) \int_{-1}^{+1} dx \,
  \frac{1}{(1 - \beta'_N \, x)^2} \right\} ,
\end{eqnarray}
where $\beta_N$, $\beta'_N$, and $\beta_y$ are the magnitudes of
$\bm{\beta_N}$, $\bm{\beta'_N}$, and $\bm{\beta_y}$, which are related by
\begin{equation}
  \bm{\beta_y} \,\equiv\, \bm{\beta_N} \, \frac{1}{2} (1 + y) \;+\;
  \bm{\beta'_N} \, \frac{1}{2} (1 - y) \, .
\end{equation}
The $x$-integrals in Eq.~(\ref{eq:softa4}) can be worked out using
\begin{equation} \label{eq:softa5}
  \int_{-1}^{+1} dx \, \frac{1}{(1 - \beta \, x)^2}
  \,=\, \frac{2}{1 - \beta^2} \, ,
\end{equation}
which yields for Eq.~(\ref{eq:softa4}) the result~:
\begin{equation} \label{eq:softa6}
  I \,= \, \frac{2 \pi}{(E'^{\mathrm{cm}}_\gamma)^2 } \, \cdot
  \left\{ -4 \,+\, 2 \, (1 - \bm{\beta_N} \cdot \bm{\beta'_N}) \,
  \int_{-1}^{+1} dy \, \frac{1}{(1 - \beta_y^2)} \right\} .
\end{equation}
By defining the variable $v \equiv \sqrt{1 + \frac{4 M_N ^2}{-t} }$, with
$t = (k - p_M)^2 = (p'_N - p_N)^2$, we express
\begin{equation} \label{eq:softa7}
  (1 - \bm{\beta_N} \cdot \bm{\beta'_N}) \,=\,
  (1 - \beta_N^2)^{1/2} \,\cdot\, (1 - {\beta'_N}^2)^{1/2} \,\cdot\,
  \left( \frac{v^2 + 1}{v^2 - 1} \right) ,
\end{equation}
and after a little algebra we can work out the remaining integral using~:
\begin{equation}
\label{eq:softa8} \int_{-1}^{+1} dy \, \frac{1}{(1 - \beta_y^2 )} \,=\,
\frac{1}{(1 - \beta_N^2)^{1/2} \,\cdot\, (1 - \beta'^2_N)^{1/2}} \,\cdot \,
\left( \frac{v^2 - 1}{2 v} \right) \, 2 \, \ln \left( \frac{v + 1}{v - 1}
\right) .
\end{equation}
Combining Eqs.~(\ref{eq:softa6}-\ref{eq:softa8}), we obtain the result for
the photon angular integral $I$,
\begin{equation}
\label{eq:softa9}
I \,= \, \frac{2 \pi}{(E'^{\mathrm{cm}}_\gamma)^2 } \, \cdot 4 \cdot
\left\{ -1 \,+\,
 \left( \frac{v^2 + 1}{2 v} \right) \, \ln \left( \frac{v + 1}{v - 1} \right)
\right\} .
\end{equation}
Using Eq.~(\ref{eq:softa9}), the soft photon limit for the cross section of
Eq.~(\ref{eq:softa2}) then leads to the result of Eq.~(\ref{eq:soft}).


\begin{thebibliography}{00}
%
\bibitem{DeTar:1988kn}
C.~DeTar and T.~Kunihiro, %
Phys. Rev. D {\bf 39} (1989) 2805.
%
\bibitem{Jido:2001nt}
D.~Jido, M.~Oka and A.~Hosaka, %
Prog. Theor. Phys. {\bf 106} (2001) 873.
%
\bibitem{Sasaki02}
S. Sasaki, T. Blum and S. Ohta, %
Phys. Rev. D {\bf 65} (2002) 074503.
%
\bibitem{Gock02}
M. G\"ockeler {\it et al.}, %
Phys. Lett. B {\bf 532} (2002) 63.
%
\bibitem{Mel02}
W. Melnitchouk {\it et al.}, %
hep-lat/0202022.
%
\bibitem{Thomas02}
A.W. Thomas, %
in Proceedings of the {\it 20th International Symposium On Lattice Field
Theory} (LATTICE 2002), to be published in Nucl. Phys. {\bf B}, Proceedings
Supplement; hep-lat/0208023.
%
\bibitem{Mor92}
G. Morpurgo, %
Phys. Rev. D {\bf 46} (1992) 4068.
%
\bibitem{Cheng02}
T.P. Cheng and L.-F. Li, %
Phys. Rev. Lett. {\bf 80} (1998) 2789.
%
\bibitem{DM02}
G. Dillon and G. Morpurgo, %
hep-ph/0211256.
%
\bibitem{Leinweber99}
D.B.~Leinweber, D.H.~Lu and A.W.~Thomas, %
Phys. Rev. D {\bf 60} (1999) 034014.
%
\bibitem{Leinweber01}
D.B. Leinweber, A.W. Thomas, and R.D. Young, %
Phys. Rev. Lett. {\bf 86} (2001) 5011.
%
\bibitem{Leinweber92}
D.B. Leinweber, T. Draper, and R.M. Woloshyn, %
Phys. Rev. D {\bf 46} (1992) 3067.
%
\bibitem{Cloet02}
I.C. Cloet, D.B. Leinweber, and A.W. Thomas, hep-lat/0302008.
%
\bibitem{Kon68}
L.A. Kondratyuk und L.A. Ponomarov, %
Yad. Fiz. {\bf 7} (1968) 11 [Sov. J. Nucl. Phys. {\bf 7} (1968) 82].
%
\bibitem{Nef78}
B.M.K. Nefkens {\it et al.}, %
Phys. Rev. D {\bf 18} (1978) 3911.
%
\bibitem{Bos91}
A. Bosshard {\it et al.}, %
Phys. Rev. D {\bf 44} (1991) 1962.
%
\bibitem{Hagiwara:2002pw}
K. Hagiwara {\it et al.} (Particle Data Group), %
Phys. Rev. D {\bf 66} (2002) 010001.
%
\bibitem{Dre83}
M.M. Giannini, %
in {\it Proceedings of the Workshop Perspectives on Nuclear Physics at
Intermediate Energies}, ICTP Trieste, Italy, 1983 (World Scientific,
Singapore, 1984); %
and D. Drechsel, in MAMI funding proposal to DFG, SFB 201 (1984-86), p. 56.
%
\bibitem{Mac99}
A.I. Machavariani, A. Faessler, and A.J. Buchmann, %
Nucl. Phys. {\bf A646} (1999) 231; %
Nucl. Phys. {\bf A686} (2001) 601 (E).
%
\bibitem{DVGS}
D. Drechsel, M. Vanderhaeghen, M.M. Giannini, and E. Santopinto, %
Phys. Lett. B {\bf 484} (2000) 236.
%
\bibitem{Drechsel:2001qu}
D. Drechsel and M. Vanderhaeghen, %
Phys. Rev. C {\bf 64} (2001) 065202.
%
\bibitem{Mach02}
A.I. Machavariani and A. Faessler, %
nucl-th/0202060.
%
\bibitem{Kot99}
M. Kotulla {\it et al.} (A2/TAPS Collaboration), %
Phys. Rev. Lett. {\bf 89} (2002) 272001.
%
\bibitem{CB}
MAMI experiment 2002, %
spokespersons R. Beck and B. Nefkens.
%
\bibitem{KSW95}
N. Kaiser, P.B. Siegel, and W. Weise, %
Phys. Lett. B {\bf 362} (1995) 23.
%
\bibitem{Isgur:1977ef}
N. Isgur and G. Karl, %
Phys. Lett. B {\bf 72} (1977) 109.
%
\bibitem{Hey:1975nc}
A.J.G. Hey, P.J. Litchfield, and R.J. Cashmore, %
Nucl. Phys. {\bf B95} (1975) 516.
%
\bibitem{Genova}
M.M. Giannini and E. Santopinto, %
private communication.
%
\bibitem{Arima:wy}
M.~Arima, K.~Shimizu, and K.~Yazaki, %
Nucl. Phys. {\bf A543} (1992) 613.
%
\bibitem{KWW97}
N. Kaiser, T. Waas, and W. Weise, %
Nucl. Phys. {\bf A612} (1997) 297.
%
\bibitem{Nieves01}
J. Nieves and E. Ruiz Arriola, %
Phys. Rev. D {\bf 64} (2001) 116008.
%
\bibitem{Inoue:2001ip}
T.~Inoue, E.~Oset and M.J.~Vicente Vacas, %
Phys. Rev. C {\bf 65} (2002) 035204.
%
\bibitem{Jido02}
D. Jido, A. Hosaka, J.C. Nacher, E. Oset, and A. Ramos, %
Phys. Rev. C {\bf 66} (2002) 025203.
%
\bibitem{ElAmiri:1992xa}
M. El Amiri, G. L\'{o}pez Castro, and J. Pestieau, %
Nucl. Phys. {\bf A543} (1992) 673.
%
\bibitem{LopezCastro:2000ep}
G. L\'{o}pez Castro and A. Mariano, %
Nucl. Phys. {\bf A697} (2002) 440.
%
\bibitem{Krusche:1995nv}
B.~Krusche {\it et al.}, %
Phys. Rev. Lett. {\bf 74} (1995) 3736.
%
\bibitem{Renard:2000iv}
F. Renard {\it et al.} (GRAAL Collaboration), %
Phys. Lett. B {\bf 528} (2002) 215.
%
\bibitem{Chiang:2001as}
W.-T. Chiang, S.N. Yang, L. Tiator, and D. Drechsel, %
Nucl. Phys. {\bf A700} (2002) 429.

\end{thebibliography}
\end{document}